\documentclass[%
preprint,
nobibnotes,
 amsmath,amssymb,
 aps,
pra,
]{revtex4-1}

\usepackage{graphicx}
\usepackage{dcolumn}
\usepackage{bm}
\usepackage{hyperref}
\usepackage{bigints}
\usepackage{booktabs}
\usepackage{epstopdf}
\usepackage{color}
\usepackage{multirow}

\begin{document}

\preprint{PKP,BB-FT}

\title{Equilibrium free energy differences at different temperatures from a single set of nonequilibrium transitions}

\author{Puneet Kumar Patra}
 \affiliation{Advanced Technology Development Center, Indian Institute of Technology Kharagpur}
\author{Baidurya Bhattacharya}%
\affiliation{Department of Civil Engineering, Indian Institute of Technology Kharagpur}
\email{baidurya@civil.iitkgp.ernet.in}

\date{\today}

\begin{abstract}
Crook's Fluctuation Theorem (CFT) and Jarzynski equality (JE) are effective tools for obtaining free energy difference $\Delta F (\lambda_A \rightarrow \lambda_B, T_0)$ through a set of finite-time  protocol driven non-equilibrium transitions between two equilibrium states $A$ and $B$ (parameterized by the time-varying protocol $\lambda(t)$) at the same temperature $T_0$. Using a new work function $\Delta W_G$, we generalize CFT to transitions between two non-equilibrium steady states (NESSs) created by a thermal gradient and show that it is possible, using the \textit{same} set of finite time transitions between these two NESSs, to obtain $\Delta F (\lambda_A \rightarrow \lambda_B, T_0)$ for different values of $T_0$, thus completely eliminating the need to make new samples for each new $T_0$. The generalized form of JE arises naturally as the average of the exponentiated $\Delta W_G$. 
The results are demonstrated on two test cases: (i) a single particle quartic oscillator having a known closed form $\Delta F$, and (ii) a 1-D $\phi^4$ chain.  Both systems are sampled from the canonical distribution at an arbitrary $T'$ with $\lambda=\lambda_A$, then subjecting it to a temperature gradient between its ends, and after steady state is reached,  effecting the protocol change $\lambda_A \rightarrow \lambda_B$ in time $\tau$, following which $\Delta W_G$ is computed.  The reverse path likewise initiates in equilibrium at $T'$ with $\lambda=\lambda_B$ and the protocol is time-reversed leading to $\lambda=\lambda_A$ and the reverse $\Delta W_G$.  Our method is found to be more efficient than either JE or CFT when free-energy differences at multiple $T_0$'s are required for the same system.
\end{abstract}

\pacs{Valid PACS appear here}
\maketitle
Consider a thermo-mechanical system whose equilibrium state is defined by its temperature $T_0$ and an external protocol $\lambda$ fixed at $\lambda_A$ (for example the position of confining potential \cite{mondaini_14}, the position of the last molecule of a protein chain \cite{No19} etc.). A large class of problems in biological and chemical physics (such as transition between conformations of proteins, folding and unfolding of proteins, enzyme-ligand binding, hydration etc.) concerns the change in free energy, $\Delta F (\lambda_A \rightarrow \lambda_B, T_0)$, of this system as its configurational space evolves under $\lambda(t)$ in a finite time $\tau$ corresponding to the final value $\lambda=\lambda_B$ and the system eventually relaxes to a new equilibrium at the same temperature $T_0$. Several methods have been proposed for computing $\Delta F (\lambda_A \rightarrow \lambda_B, T_0)$ - thermodynamic integration \cite{kirkwood_35}, umbrella sampling \cite{torrie_77}, steered molecular dynamics \cite{park_03}, and nonequilibrium work relations \cite{jarzynski_97a, jarzynski_97b, jarzynski_07, crooks_98, No5, hatano_99}. 

The development of Jarzysnki's equality (JE) \cite{jarzynski_97a, jarzynski_97b,jarzynski_07} and Crooks' fluctuation theorem (CFT) \cite{crooks_98,No5,No6} has dramatically improved our ability to  calculate free-energy differences \citep{No14,No15,No16,No17} of real systems \citep{No18,No19} through finite-time irreversible processes between two equilibrium states at the same temperature, $T_0$. Nevertheless, the task remains daunting because of the requirement of extensive sampling of the configurational space. In addition, $\Delta F (\lambda_A \rightarrow \lambda_B, T_0)$ thus computed is valid only for the particular temperature at which the samplings are performed and if $\Delta F (\lambda_A \rightarrow \lambda_B, T^\prime \neq T_0)$, is needed, the re-sampling of the entire data set is necessary at $T^\prime$.

In this work we generalize CFT and JE by proposing a new fluctuation theorem that enables us to calculate $\Delta F (\lambda_A \rightarrow \lambda_B, T_0)$ with good accuracy for a range of $T_0$ values using \textit{a single set of sampling data}, thereby completely eliminating the need to make new samples for each new $T_0$. The proposed fluctuation theorem utilizes the transition between two nonequilibrium states, and $T_0$ features in the equation as a scaling parameter.  

Let us now look at details of the problem. For a system in canonical equilibrium, the Helmholtz free energy is:
\begin{equation}
F(T_0,\lambda) = -k_BT_0 \log \left( \int \exp \left[-\beta_0 E(\Gamma,\lambda) \right] d\Gamma \right),
\label{eq:one}
\end{equation}
where $k_B$ is the Boltzmann constant and $\beta_0 = (k_BT_0)^{-1}$. The system's energy $E(\Gamma,\lambda)$ depends upon the microstate $\Gamma$ and varies parametrically over time according to $E(\Gamma(t),\lambda(t)) = \sum p_i^2/2m + \Phi(x_1,x_2,\ldots,x_N,\lambda(t))$, where $p_i$ and $x_i$ are the momentum and the position of the $i^{\text{th}}$ particle. In CFT, the system is initially in equilibrium state $A$ with $\lambda = \lambda_A$. At time $t=0$, $\lambda$ starts to evolve until $t=\tau$, and stays fixed at its new value $\lambda_B$. During this period work $W^F = \int_0^\tau \dot{\lambda} \partial E / \partial \lambda  dt$ is performed. The superscript $F$ denotes the forward transition $A \rightarrow B$. Over time, the system relaxes to a new equilibrium state $B$ with $\lambda=\lambda_B$. Being irreversible, the work $W^F$ depends upon the initial microstate $\Gamma(0)$ of the system (and its surroundings), and therefore, an exhaustive sampling of the initial microstates provides the probability density of forward work, $P(W^F=w)$. Now consider the same system evolving in a reverse manner. The system begins at equilibrium state $B$ where $\lambda = \lambda_B$, and over $0 \leq t \leq \tau$, $\lambda$ traces itself back from $\lambda_B \rightarrow \lambda_A$. Eventually the system reaches the equilibrium state $A$. Repeated sampling of this reverse transition provides $P(W^R=-w)$ for the reverse work. CFT relates the ratio of these two densities with $\Delta F (\lambda_A \rightarrow \lambda_B, T_0)$:
\begin{equation}
\frac{P(W^{F} = w_0)}{P(W^{R} = -w_0)} = \exp[-\beta_0 (w_0 - \Delta F(T_0,\lambda))]
\label{eq:cft}
\end{equation}
The validity of \ref{eq:cft} requires the dynamics to be ergodically consistent i.e. {if a microstate has a nonzero probability in equilibrium state $A$, it evolves to a microstate that has a nonzero probability in equilibrium state $B$.} Integrating (\ref{eq:cft}) gives JE \citep{No11}. However, since $T_0$ is implicit in the sampling dynamics, the probability densities obtained cannot be used to calculate $\Delta F(T^\prime, \lambda)$ if $T^\prime \neq T_0$. In order to employ a single set of sampling data for calculating $\Delta F(T^\prime, \lambda)$ corresponding to a range of temperature $T^\prime$, the dependence of sampling data on $T_0$ must be removed. We set out to do this by looking at the work and heat distributions during the \textit{transition between two nonequilibrium steady states}.

Rather than beginning at equilibrium, we begin at a nonequilibrium steady state $SS_1$ obtained by imposing a temperature difference $(T_H - T_C)$ at the two ends of the conductor, where $T_H$ and $T_C$ are the temperatures of the hot and cold ends. {This steady state originated from some primordial arbitrary equilibrium state $A$ characterized by $\lambda_A$ and $T_0$ by employing suitable temperature constraints.} {For all practical purposes, the system reaches steady-state when the relevant time-averaged macroscopic observables become stationary.} $T_H, T_C$ and $T_0$ are related to each other through $T_H = T_0 + \Delta T_H$ and $T_C = T_0 - \Delta T_C$. Thus, depending upon $\Delta T_H$ and $\Delta T_C$, {both not necessarily being equal,} one can think of starting from arbitrarily different canonical equilibrium states. {Note that this allows us to choose any arbitrary $T_0$.}

After $SS_1$ is achieved, at $t=0$, $\lambda$ starts to evolve from $\lambda_A$ until time $t=\tau$ when $\lambda = \lambda_B$ and work is performed. {This external work does not result in any phase-space compression.} Given sufficient time, the system reaches a new steady state $SS_2$. Upon removing the temperature constraints, the system eventually reaches the equilibrium state $B$, defined by $\lambda_B$ and $T_0$. The reverse transition can likewise be accomplished under the time-reversed protocol. Such transition between steady-states has been studied before in a different context\cite{lahiri_14}. The underlying principle governing our approach is the relaxation of a nonequilibrium state to an equilibrium state \citep{No23}. This relaxation is governed by the constraints imposed on the system, and thus one can obtain a multitude of equilibrium states from a single nonequilibrium state by judiciously choosing the constraints and boundary conditions. 

In state $A$ (state $B$), the system follows the canonical distribution parameterized by $\lambda_A$ (by $\lambda_B$):
\begin{equation}
f_{eq,A}(\Gamma) = \dfrac{1}{Z_{\lambda_A}}\exp \left[ -\beta_0 E(\Gamma,\lambda_A) \right]
\label{eq:canonical}
\end{equation}
The density function of the nonequilibrium state and the Jacobian are given by Liouville's equation \cite{No21}:
\begin{equation}
\begin{array}{rcl}
f_{SS,1} \left[\Gamma(t)\right] & = & f_{eq,A} \left[\Gamma(0)\right] \exp \left[ - \int\limits_0^t dt^\prime \left( \Lambda_H(t^\prime) + \Lambda_C(t^\prime) \right) \right], \\
d\Gamma(t) & = &d\Gamma(0) \exp \left[ \int\limits_0^t dt^\prime \left( \Lambda_H(t^\prime) + \Lambda_C(t^\prime) \right) \right],
\end{array}
\label{eq:steady_state}
\end{equation}
where {$\Lambda = \left[ \partial \dot{\Gamma}/ \partial \Gamma\right]$ denotes the phase-space compression factor, with $H$ ($C$) denoting the hot (cold) region. The intermediate region does not contribute to $\Lambda$ (owing to Hamilton's equation of motion).} Importantly, the normalizing constant corresponding to $SS_1$ is the same as the partition function for $A$. The phase-space compression factors are related to the heat flow \cite{bright2005new,higherBT,heat_pump,patra2015equivalence} from the thermostats through:
\begin{equation}
\begin{array}{rcl}
\langle \dot{Q}_H \rangle_t = k_B T_H \langle \Lambda_H \rangle _t t &,& \langle \dot{Q}_C \rangle_t = k_B T_C \langle \Lambda_C \rangle _t t.
\end{array}
\label{eq:heat_flow_ph_comp}
\end{equation}
{For sake of compactness, we will drop $t$ from the density functions and cumulative heat flows later.} Next, we bring the generalized dimensionless time-integrated work function, $\Delta W_G(t)$ \cite{No25} into picture, which can relate {two microstates ($\Gamma(0)$ and $\Gamma(t)$), neither of them necessarily in equilibrium:}
\begin{equation}
\exp \left( \Delta W_G (t) \right) = \dfrac{f_1(\Gamma(0)) d\Gamma(0) Z_{\lambda_0} }{f_2(\Gamma(t)) d\Gamma(t) Z_{\lambda_t}},
\label{eq:gen_work_func}
\end{equation} 
{The initial microstate $\Gamma(0)$ evolves to $\Gamma(t)$ in time $t$. $f_1(\Gamma(0))$ (or $f_2(\Gamma(t))$) is the probability density of $\Gamma(0)$ (or $\Gamma(t)$) corresponding to an associated equilibrium state 1 (or 2)}. We conjecture that such an association is possible after the system undergoing non-equilibrium transition loses its memory.  $Z_{\lambda_i}$ denotes the partition function at $\lambda_i$. Now we bring the superscripts $F$ (for the forward transition $A \to SS_1 \to SS_2 \to B$) and $R$ (for the reverse transition $B \to SS_2 \to SS_1 \to A$). The forward transition takes $\Gamma(0) \to \Gamma(t)$, while the reverse transition takes $\Gamma^\ast(0) \to \Gamma^\ast(t)$, where $\Gamma^\ast(0)$ is related to $\Gamma(t)$ through time-reversal mapping. The generalized work function during $A \to SS_1$ is {(see Section-I of Appendix)}:
\begin{equation}
\Delta W^F_{G,A\to SS_1} = \frac{1}{T_0}\int_0^t\left[\dfrac{\Delta T_H}{T_H} \dot{Q}_{H}^F - \dfrac{\Delta T_C }{T_C}\dot{Q}_{C}^F\right]dt^\prime
\label{eq:work_a_to_SS1}
\end{equation}
Proceeding analogously {(see Section-III of Appendix)}, the generalized work function during $A \to SS_2$ is:
\begin{equation}
\Delta W^F_{G,A\to SS_2} = \frac{1}{T_0}\int_0^{t+\tau}\left[\dfrac{\Delta T_H}{T_H} \dot{Q}_{H}^F - \dfrac{\Delta T_C }{T_C}\dot{Q}_{C}^F + \dot{W}^F \right]dt^\prime
\label{eq:work_a_to_SS2}
\end{equation}
Therefore, the work function during $SS_1 \to SS_2$ can be obtained by subtracting \ref{eq:work_a_to_SS1} from \ref{eq:work_a_to_SS2}:
\begin{equation}
\Delta W_{G,SS_1 \to SS_2}^F  = \beta_0 W^{F} + \frac{1}{T_0} \left[\dfrac{\Delta T_H}{T_H} {Q}_{H,\tau}^F - \dfrac{\Delta T_C}{T_C} {Q}_{C,\tau}^F\right],
\label{eq:work_ss1_to_ss2}
\end{equation} 
where the heat flows are for the time duration $\tau$ over which $\lambda$ changes. In a similar manner, we can compute the work function during the reverse transition $SS_2 \to SS_1$:
\begin{equation}
\Delta W_{G,SS_2 \to SS_1}^R = \beta_0W^{R} + \frac{1}{T_0} \left[\dfrac{\Delta T_H}{T_H} Q_{H,\tau}^R - \dfrac{\Delta T_C}{T_C} Q_{C,\tau}^R\right],
\label{eq:eight_seventeen}
\end{equation} 

\begin{figure}
\small
\centering
\includegraphics[scale=0.65]{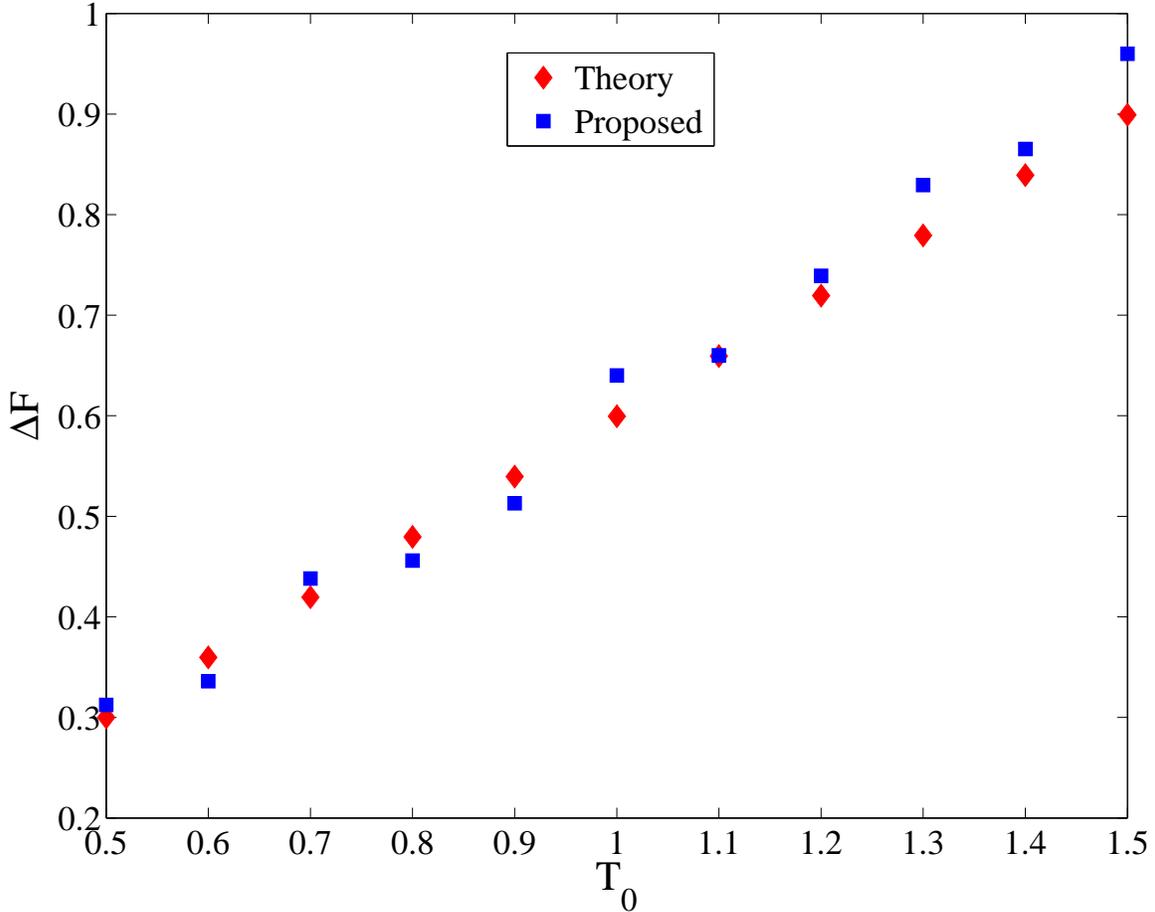}
\caption{Comparison of $\Delta F$ obtained using theoretical and proposed approaches in test case 1. Notice, that the proposed approach provides a good approximation to the theoretical results.}
\label{fig:figure_single}
\end{figure}

Now we make the important assumption of the ergodic consistency being valid during the transition $SS_1 \to SS_2$, and therefore, using \ref{eq:gen_work_func} we can write: 
\begin{equation}
\begin{array}{ccc}
\exp \left[ \Delta W_{G,SS_1 \to SS_2}^F \right]  & = & \dfrac{ f_{eq,A} \left[ \Gamma_{SS,1}(0)\right] d\Gamma_{SS,1}(0) Z_{\lambda_A}}{f_{eq,B} \left[ \Gamma_{SS,2}(\tau)\right] d\Gamma_{SS,2}(\tau) Z_{\lambda_B}} \\
\end{array}
\label{eq:relation}
\end{equation} 
The subscripts $SS_i$ emphasize that the points are on trajectories whose evolution is described by equations of motion that take the ensemble of states from $SS_1$ at 0 to $SS_2$ at $\tau$. Because of the deterministic nature of the dynamics, $\Delta W_{G,SS_1 \to SS_2}^F=-\Delta W_{G,SS_2 \to SS_1}^R$. For simplicity, we now drop all subscripts except $G$. The probability densities of the forward and reverse work functions therefore can be related as {(see Section-II of Appendix)}:
\begin{equation}
\begin{array}{rcl}
P \left[ \Delta W_G^R = -k \right] =  e^{-k} \dfrac{Z_{\lambda_A}}{Z_{\lambda_B}} P \left[ \Delta W_G^F = k \right]
\label{eq:proposed_ft_first}
\end{array}
\end{equation}
A rearrangement results in the \textit{proposed fluctuation relation}:
\begin{equation}
\frac{P \left[\beta_0W^F + \beta_0 \left[\dfrac{\Delta T_H}{T_H}{Q}_{H}^F - \dfrac{\Delta T_C}{T_C} {Q}_{C}^F\right] = k\right]}{P \left[\beta_0W^R + \beta_0\left[\dfrac{\Delta T_H}{T_H} {Q}_{H}^R - \dfrac{\Delta T_C}{T_C}{Q}_{C}^R\right] = -k\right]} = e^{\left[k - \beta_0\Delta F \right]},
\label{eq:proposed_ft}
\end{equation}
which is the main result of this paper (henceforth, referred to as GCFT). Since the samplings have been performed at $T_H$ and $T_C$, the effect of $T_0$ is \textit{inherently absent} in them, and $\beta_0$ is simply a scaling parameter. Depending upon the temperature at which $\Delta F$ is to be calculated, we can compute the forward and reverse densities of the work function simply by substituting the desired value of $T_0$. The generalized JE may be obtained by averaging: 
\begin{equation}
\left\langle e^{\left[-\beta_0W^{F} - \beta_0\left[\dfrac{\Delta T_H}{T_H} {Q}_{H}^F - \dfrac{\Delta T_C}{T_C} {Q}_{C}^F \right] \right]} \right\rangle = e^{\left[ - \beta_0\Delta F \right]}.
\label{eq:gje}
\end{equation}
A second law type inequality can be recovered by applying the Jensen's inequality to (\ref{eq:gje}):
\begin{equation}
\begin{array}{rcl}
\langle W^{F} \rangle + \langle {Q}_H^F \rangle \dfrac{\Delta T_H}{T_H} - \langle {Q}_C^F \rangle \dfrac{\Delta T_C}{T_C}   & \geq & \Delta F \\
\end{array}
\label{eq:eight_twentysix}
\end{equation}
\textcolor{black}{It must be noted that the above equations are \textit{not} exact relationships, and hold true \textit{only} for large $\tau$.} Taking $\tau$ large enough, while fixing the time required to reach the steady state, ensures that the contributions arising from phase-space compressions become negligible. We test the effectiveness of (\ref{eq:proposed_ft}) on a 1-D $\phi^4$ chain of $N$ particles. Its energy function is:
\begin{equation}
\begin{array}{ccc}
E & = & \sum \limits_{i=1}^N \dfrac{p_i^2}{2m_i} + \sum\limits_{i=1}^NU\left(x_i,x_{i+1} \right) + \sum\limits_{i=1}^N V(x_i)
\end{array}.
\label{eq:seven_one}
\end{equation}
Here $U(x_i,x_{i+1}) = 0.5k_1 \left( | x_{i+1} - x_i | - d \right)^2$ represents the quadratic nearest neighbour interparticle interaction, while $V(x_i) = 0.25 k_2 \left( x_i - x_{i,0} \right)^4$ represents the quartic tethering potential with $x_{i,0}$ being the equilibrium position of the $i^{th}$ particle. We have kept $k_1= 1.0$ and $m_i = 1.0$. $k_2$ plays the role of $\lambda$: 
\begin{equation}
\begin{array}{ccll}
k_2 & = & 0.25 & \forall t < 0 \\
& = & 0.25 \left( 1 + {10 t}/{\tau} \right) & \forall 0 < t \leq \tau \\
& = & 2.75 & \forall t > \tau 
\end{array}
\label{eq:eight_twentytwo}
\end{equation}
{\textit{Test Case 1:} The first test case involves a single particle system (subscript 1 dropped) having a known analytical solution for $\Delta F$:
\begin{equation}
\begin{array}{rcl}
\Delta F & = & -k_BT_0\log\left( \dfrac{\int e^{\left[ -\beta_0 11x^4/4 \right]} e^{\left[ -\beta_0 p^2/2 \right]} dx dp}{\int e^{\left[ -\beta_0 x^4/4 \right]}e^{\left[ -\beta_0 p^2/2 \right]} dx dp} \right)\\
& = & k_BT_0/4 \log(11)\\
\end{array}
\label{eq:new_example1}
\end{equation}
We compare this known $\Delta F$ with our results. We subject the single quartic oscillator to a position-dependent temperature field,
\begin{equation}
T(x) = 1 + 0.1\tanh(x),
\label{eq:new_example2}
\end{equation}
to bring it away from equilibrium. Temperature is controlled by Hoover-Holian thermostat \cite{Hoover1996253}. The system is simulated for 100,000 time steps (each time step = 0.001) under this temperature field through which it reaches $SS_1$. $k_2$ is changed over the next $\tau = $ 10,000 time steps according to equation (\ref{eq:eight_twentytwo}). $\Delta W_G$ in this case is:
\begin{equation}
\Delta W_G = \beta_0 W + \beta_0 \int_0^{\tau}\dot{Q}dt^\prime - \int_0^{\tau}\beta(x)\dot{Q}dt^\prime,
\label{eq:new_example3}
\end{equation}
where, $\beta(x) = 1/k_BT(x)$ and $\dot{Q} = -\eta T(x) - 3p^2\xi T(x)$. $\eta$ and $\xi$ are the Hoover-Holian thermostat variables. $W=\mathcal{H}(\tau) - \mathcal{H}(0)$, where $\mathcal{H}(t) = p^2/2 + V + \int(\eta T(x) + 3p^2\xi T(x))dt^\prime$. Probability densities of generalized work are constructed using 60,000 random initial points. Figure \ref{fig:figure_single} shows $\Delta F$ due to the evolution of $k_2$ as a function of temperature: GCFT is able to reproduce the theoretical results accurately for a range of temperatures without the need to resample at every new $T_0$.}

\begin{figure}
\small
\centering
\includegraphics[scale=0.65]{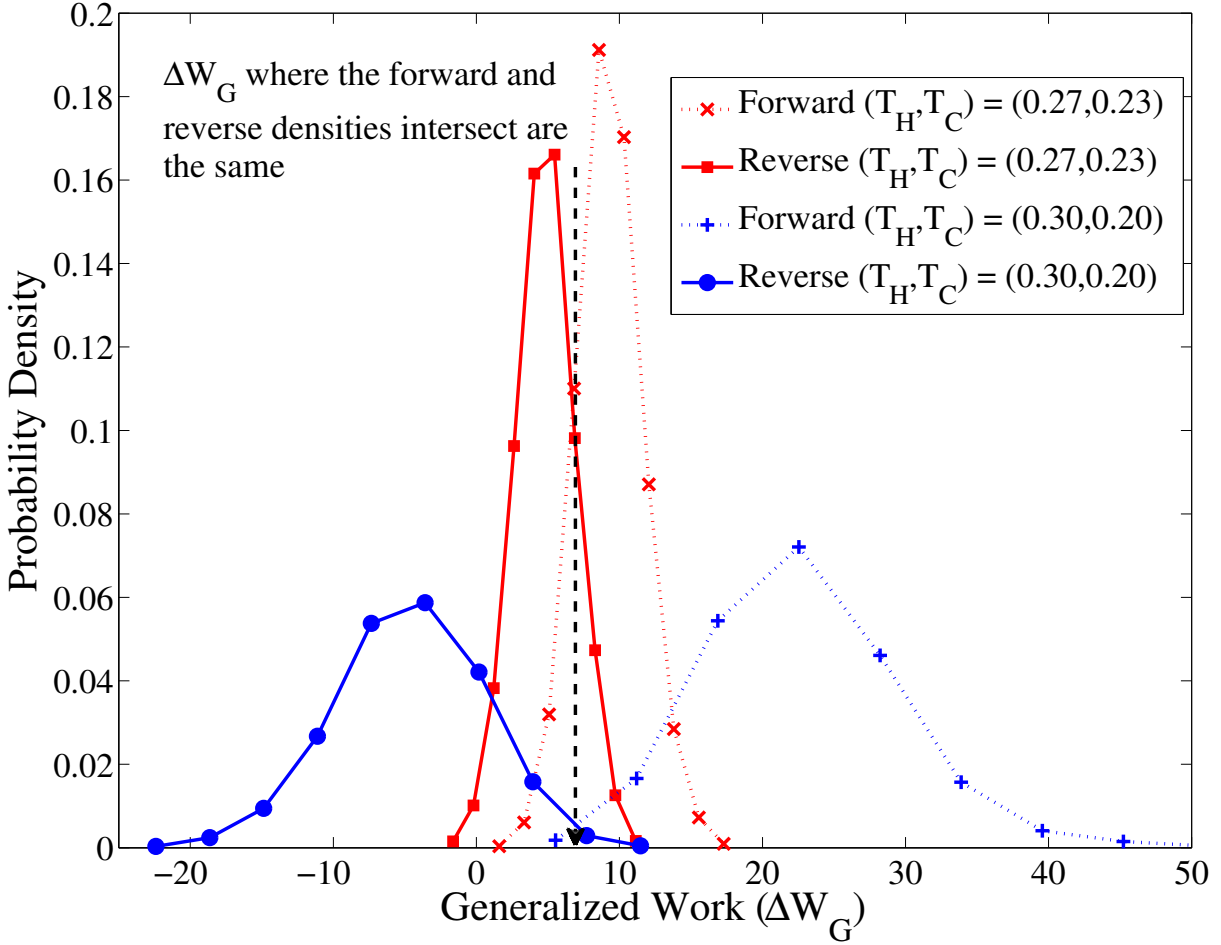}
\caption[Forward and reverse probabilities of work for CFT]{\label{fig:8_one} Forward and reverse probabilities of generalized work function at $T_0 = 0.29$  calculated using 5,000 forward and reverse trajectories (red) ($T_H,T_C$) = (0.27,0.23) and (green) ($T_H,T_C$) = (0.30,0.20). The forward and reverse probabilities at approximately the same value of $W_G$. $\Delta F$ calculated compares well with that out JE and CFT. Results obtained using the same dataset for other $T_0$ values are similar, and agree well with CFT.}
\end{figure}

{\textit{Test Case 2:} We now consider a larger system ($N = 25$). The system is initialized with $x_i=x_{i,0}=i$ and random particle velocities. The equations are integrated using classic Runge-Kutta algorithm with an incremental time step of 0.01. Post initialization, a temperature gradient is imposed on the system by keeping the two end particles at $T_H$ and $T_C$ using two Nos\'e-Hoover (NH) thermostats \cite{nose_hoover}. Subsequently, after 1 million timesteps (steady state is assumed to have reached), $k_2$ evolves in $\tau = 100,000$ time steps. The cumulative heat flow from the hot thermostat is $Q_{H} = -\int_{0}^{t} T_H \eta_{H}dt^\prime$ (likewise for the cold), where $\eta_H (\eta_C)$ is the hot (cold) NH variable.  The work done due to the change in tethering potential during time $\tau$ is  $W=\mathcal{H}(\tau) - \mathcal{H}(0)$ where
\begin{equation}
\begin{array}{lcr}
\mathcal{{H}}(t) = \sum \dfrac{p_i^2}{2} + \Phi + \int_0^t \eta_C p_C^2 dt^\prime + \int \eta_H p_H^2 dt^\prime & & \\
\end{array}
\label{eq:eight_twentythree}
\end{equation}
Here $p_{H}$ ($p_C)$ denotes the hot (cold) particle's momentum, and $\Phi = \sum U + \sum V$. $\Delta W_G^F$ and $\Delta W_G^R$ are computed using 5,000 trajectories each. Figure \ref{fig:8_one} shows probability densities of the forward and reverse generalized work functions $P(\Delta W_G^F)$ and $P(\Delta W_G^R)$ at $T_0 = 0.29$. Two pairs of $(T_H,T_C)$ - red for (0.27,0.23) and blue for $(0.30,0.20)$ - are chosen. The points of intersection of the forward-reverse pair gives $\beta_0 \Delta F$ (\ref{eq:proposed_ft}) which should be independent of $(T_H,T_C)$ for the same $T_0$ as evident from the figure. Importantly, these same 10000 samples can be used to compute $\Delta F$ at \textit{any} $T_0$. Table (\ref{tab:8one}) shows seven such $T_0$ values, computed using both sets of  $(T_H,T_C)$. Not only is $\Delta F$ at a given $T_0$ independent of $(T_H,T_C)$ as it should be, it is clear that $T_0$ does not even need to be within the range of $(T_H,T_C)$ for the method to work.} 
 
{Finally, Table (\ref{tab:8one})lists $\Delta F$ computed using JE and CFT at the seven different temperatures. While GCFT is able to identify $\Delta F$ as accurately as CFT and JE, it does so with only one set of samples. CFT and JE on the other hand would require a new set of samples for each $T_0$, thereby imposing a severe computational or experimental burden on the analyst. We must, however, point out that the transition needs to be carried out slowly, as our efforts to calculate $\Delta F$ using $\tau = 100$ steps did not yield any fruitful result.}

\begin{table} 
\small
\centering
\caption[Equilibrium free energy difference]{\label{tab:8one}Comparison of free energy differences using JE, CFT and GCFT for seven different values of $T_0$. GCFT results are for two different steady-states: $T_H=0.27,T_C=0.23$ and $T_H=0.30,T_C=0.20$. Notice that the $\Delta F$ obtained using GCFT matches closely with those from JE and CFT. It is interesting to note that the case of $T_H=0.27,T_C=0.23$ is able to approximate the equilibrium free energy differences even for the states as far as $T_0=0.21$ and $T_0=0.29$. The results indicate that one can use a single set of data obtained during a transition between two NESS and employ GCFT to calculate free energy differences for a range of temperature. }
\begin{tabular}{|c|c|c|c|c|}
\multirow{2}{*}{$T_0$} & \multirow{2}{*}{JE} & \multirow{2}{*}{CFT} & GCFT                        & GCFT                \\ 
                       &                     &                      & $(T_H,T_C) = (0.30,0.20)$ & $(T_H,T_C)=(0.27,0.23)$ \\ \hline \hline
0.21                   & 1.39               & 1.35                & 1.40                       & 1.39               \\
0.22                   & 1.48               & 1.50                & 1.49                       & 1.46               \\
0.24                   & 1.64               & 1.60                & 1.67                       & 1.62               \\
0.25                   & 1.73               & 1.70                & 1.75                       & 1.69               \\
0.26                   & 1.81               & 1.79                & 1.83                       & 1.78               \\
0.28                   & 1.98               & 1.97                & 2.02                       & 1.97               \\
0.29                   & 2.07               & 2.00                & 2.10                       & 2.07           \\ \hline \hline   
\end{tabular}
\end{table}

\textit{To summarize} in this work, generalized versions of CFT and JE have been presented. The proposed extensions present a suitable method through which equilibrium free energy differences can be extracted from the information embedded within the non-equilibrium steady states. The augmented equations bear remarkable similarity with those of CFT and JE with additional contributions arising due to heat flowing from the reservoirs. GCFT has been tested using two different cases, with each of them suggesting that GCFT is a suitable alternative to CFT and JE when evaluating $\Delta F$ at multiple temperatures.

\section{Appendix}
\subsection{Section-I}
In this section we will derive equation (7) of the manuscript. Let us look at a system initially in canonical equilibrium (state $A$, temperature $T_0$), whose distribution function given by:
\begin{equation}
f_{eq,A}(\Gamma(0)) = \dfrac{1}{Z_A}e^{-\beta_0 E(\Gamma(0))},
\label{eq:appone}
\end{equation}
for a microstate $\Gamma(0)$ of $A$. $Z_A$ is the partition function and $\beta_0 = (k_BT_0)^{-1}$. On this system, we apply a temperature gradient by keeping the two ends at temperatures $T_H$ and $T_C$. Because of the dynamical nature the system evolves to a new microstate $\Gamma(t)$ in time $t$ under the influence of the thermal gradient. Assuming deterministic dynamics, the system evolves according to Liouville's continuity equation \cite{No20}, and the evolved distribution function becomes:
\begin{equation}
\begin{array}{rcl}
\dfrac{df}{dt} = -f\Lambda & = & -f \dfrac{\partial \dot{\Gamma}}{\partial \Gamma} \\
\implies f(\Gamma(t)) & = & f_{eq,A}(\Gamma(0))\exp\left[{-\int_0^t\Lambda dt^\prime} \right] \\
&  = & f_{eq,A}(\Gamma(0))\exp\left[ \langle \Lambda \rangle_tt \right],
\label{eq:apptwo}
\end{array}
\end{equation}
where $\langle . \rangle_t$ denotes the time-average. The term $\Lambda$ signifies the phase-space compression factor, which denotes the average rate at which the phase-space collapses onto a fractal dimension smaller than the ostensible dimension \cite{No20}. $\Lambda$ can be related to several important dynamical variables like Lyapunov exponents \cite{bright2005new}, and thermodynamic variables like heat flow ($\dot{Q}$) \cite{No22} and entropy production ($\dot{S}$) \cite{heat_pump,higherBT,patra2015equivalence}:
\begin{equation}
\dot{S} = \dfrac{\langle\dot{Q}\rangle}{T} = k_B \langle \Lambda \rangle.
\label{eq:appthree}
\end{equation}
The phase-space compresses (or expands) due to the heat flows from the individual thermostatted regions (the intermediate regions do not contribute towards phase-space compression owing to Hamilton's evolution equation), and may be split up into two parts:
\begin{equation}
\Lambda \equiv \Lambda_H +  \Lambda_C = \dfrac{\dot{Q}_H}{k_BT_H} + \dfrac{\dot{Q}_C}{k_BT_C}.
\label{eq:appfour}
\end{equation}
Utilizing equation (\ref{eq:apptwo}), we may write the nonequilibrium distribution post time $t$ as:
\begin{equation}
\begin{array}{rcl}
f_{neq}(\Gamma(t)) & = & f_{eq,A}(\Gamma(0))\exp\left[{-\int_0^t\Lambda dt^\prime} \right] \\
& = & f_{eq,A}(\Gamma(0))\exp\left[\left( \langle \Lambda_H \rangle_t + \langle \Lambda_C \rangle_t \right)t \right]
\label{eq:appfive}
\end{array}
\end{equation}
Equation (\ref{eq:appfive}) represents the general nature of a nonequilibrium distribution function, and therefore, represents a steady-state distribution function as well. Another important conclusion from equation (\ref{eq:appfive}) is:
\begin{equation}
\begin{array}{rcl}
f_{neq}(\Gamma(t))d\Gamma(t) & = & f_{eq,A}(\Gamma(0)) d\Gamma(0) \\
\implies \dfrac{d\Gamma(0)}{d\Gamma(t)} & = & \exp \left[ - \int_0^t \Lambda dt^\prime \right].
\label{eq:appsix}
\end{array}
\end{equation}
The generalized work function, $\Delta W_G$, introduced in the main text is the same as the one in Williams \textit{et. al.} \cite{No25}, and relates two microstates that are not necessarily in equilibrium:
\begin{equation}
\exp \left[ \Delta W_G \right] = \dfrac{f_1(\Gamma(0))d\Gamma(0) Z_{\lambda_0}}{f_2(\Gamma(t))d\Gamma(t) Z_{\lambda_t}}.
\label{eq:appseven}
\end{equation}
\textit{For deriving equation (7)} of the manuscript, we look at the transition between the equilibrium state $A$, and the nonequilibrium steady-state obtained after introducing the thermal gradient (for a time $t$). During this transition, $\lambda$ does not change, and as a result, the partition functions may be omitted. Here, $f_1$ denotes the canonical distribution function shown in equation (\ref{eq:appone}), and $f_2$ denotes the canonical distribution function associated with the nonequilibrium microstate $\Gamma(t)$. Appropriate substitution results in:
\begin{equation}
\exp \left[ \Delta W_G \right] = \dfrac{\exp \left[ -\beta_0 E(\Gamma(0)) \right] d\Gamma(0)}{\exp \left[ -\beta_0 E(\Gamma(t)) \right] d\Gamma(t)}.
\label{eq:appeight}
\end{equation}
The ratio of the differential volume terms are related to the phase-space compression factor:
\begin{equation}
\dfrac{d\Gamma(0)}{d\Gamma(t)} = \exp \left[ - \int_0^t \Lambda dt^\prime \right].
\label{eq:appnine}
\end{equation}
Substituting equation (\ref{eq:appnine}) in equation (\ref{eq:appeight}), we get:
\begin{equation}
\exp \left[ \Delta W_G \right] = \dfrac{\exp \left[ -\beta_0 E(\Gamma(0)) \right]}{\exp \left[ -\beta_0 E(\Gamma(t)) \right]} \times \exp \left[ - \int_0^t \Lambda dt^\prime \right].
\label{eq:appten}
\end{equation}
Employing the first law of thermodynamics, equations (\ref{eq:appthree}) and (\ref{eq:appfour}), and recognizing that no external work is performed during  the transition from $A \to SS_1$, we can write:
\begin{equation}
\begin{array}{rcl}
\dot{E}  = \dot{Q} \implies E(\Gamma(t)) & = & E(\Gamma(0)) + \int_0^t \left[ \dot{Q}_H + \dot{Q}_C \right] dt^\prime \\
& = & E(\Gamma(0)) + k_B \int_0^t \left[ T_H \langle \Lambda_H \rangle + T_C \langle \Lambda_C \rangle \right] dt^\prime.
\label{eq:appeleven}
\end{array}
\end{equation}
Substituting equation (\ref{eq:appeleven}) into equation (\ref{eq:appten}), we get:
\begin{equation}
\begin{array}{rcl}
\exp \left[\Delta W_G \right] & = & \exp \left[ -\beta_0 \left( E(\Gamma(0)) - E(\Gamma(t)) \right) \right] \exp \left[ -\int_0^t\Lambda dt^\prime \right] \\

\implies \exp \left[\Delta W_G \right] & = & \exp \left[ \beta_0 \left( \int_0^t \left( \dot{Q}_H + \dot{Q}_C \right)dt^\prime \right) \right] \exp \left[ -\int_0^t\left( \Lambda_H + \Lambda_C \right) dt^\prime \right] \\

\implies \exp \left[\Delta W_G \right] & = & \exp \left[ \beta_0 \left( \int_0^t \left( \dot{Q}_H + \dot{Q}_C \right)dt^\prime \right) \right] \exp \left[ -\dfrac{1}{k_B}\int_0^t\left( \dfrac{\dot{Q}_H}{T_H} + \dfrac{\dot{Q}_C}{T_C} \right) dt^\prime \right] \\

\implies \exp \left[\Delta W_G \right] & = & \exp \left[ \beta_0 \int_0^t \left( \dfrac{T_H-T_0}{T_H}\dot{Q}_H + \dfrac{T_C-T_0}{T_C}\dot{Q}_C\right) dt^\prime \right] \\

\implies \exp \left[\Delta W_G \right] & = & \exp \left[ \beta_0 \left( \dfrac{\Delta T_H}{T_H} Q_{H,t} - \dfrac{\Delta T_C}{T_C} Q_{C,t}\right) \right] \\
\end{array}
\label{eq:apptwelve}
\end{equation}
Since the relation (\ref{eq:apptwelve}) holds true for a generalized nonequilibrium state, it \textit{must} hold true for the nonequilibrium steady-state as well. Therefore, we write:
\begin{equation}
\Delta W^F_{G,A \to SS_1}  = \beta_0 \left( \dfrac{\Delta T_H}{T_H} Q_{H,t} - \dfrac{\Delta T_C}{T_C} Q_{C,t}\right),
\nonumber
\end{equation}
which is the same as the equation (7) of the manuscript.

\subsection{Section-II}
We will use the generalized work function to (i) derive the fluctuation theorem for heat flow \cite{evans_ft_heatflow}, and (ii) calculate the probability of violation of Fourier's law in thermal conduction \cite{No22}. Rewriting (\ref{eq:apptwelve}) in terms of the phase-space compression factors, we get:
\begin{equation}
\Delta W_G  = \beta_0 \left( \int_0^t \left[ k_B \Delta T_H \Lambda_H - k_B \Delta T_C \Lambda_C \right] dt^\prime \right),
\label{eq:appthirteen}
\end{equation}
We now take the special case that $T_0$ is the average of $T_H$ and $T_C$ \cite{evans_ft_heatflow,No22},
\begin{equation}
T_0 = \dfrac{T_H + T_C}{2}; \Delta T_H = \Delta T_C = \dfrac{T_H - T_C}{2}.
\nonumber
\end{equation}
Substituting this in equation (\ref{eq:appthirteen}), we get:
\begin{equation}
\Delta W_G  = \dfrac{1}{T_H + T_C} \int_0^t \left[ (T_H - T_C) \Lambda_H - (T_H - T_C) \Lambda_C \right]dt^\prime = \dfrac{T_H - T_C}{T_H + T_C} \int_0^t \left[ \Lambda_H - \Lambda_C \right]dt^\prime .
\label{eq:appfourteen}
\end{equation}
For a Nos\'e-Hoover thermostatted system in a \textit{d}-dimensional phase-space, having $N_T$ particles under the influence of the thermostats, the phase-space compression may be written as:
\begin{equation}
\Lambda_H = - dN_T\eta_H; \Lambda_C = -dN_T\eta_C,
\label{eq:appfifteen}
\end{equation}
where $\eta$ represents the Nos\'e-Hoover reservoir variable. Equation (\ref{eq:appfourteen}) may now be written as:
\begin{equation}
\Delta W_G^F  =  dN_T \dfrac{T_H - T_C}{T_H + T_C} \int_0^t \left[ \eta_C - \eta_H \right]dt^\prime .
\label{eq:appsixteen}
\end{equation}
We have introduced the superscript $F$ to denote the time-forward motion. Equation (\ref{eq:appsixteen}) \textit{is the same as the equation 14 derived by Evans \textit{et. al.}} \cite{No22}. It is evident that $\Delta W_G^F$ depends upon the initial microstate from which the trajectory initiates. Therefore, $\Delta W_G^F$ in equation (\ref{eq:appsixteen}) may be written as $\Delta W_G^F (\Gamma(0))$. To obtain the fluctuation theorem for heat flow \cite{evans_ft_heatflow}, we will look at the time-reversed dynamics. In the time-reversed dynamics, the system begins at the nonequilibrium microstate $\Gamma^\ast(0)$ which is the same microstate as $\Gamma(t)$ but with reversed momenta. The system reaches in time $t$ the microstate $\Gamma^\ast(t)$ which is the same microstate as $\Gamma(0)$ but with reversed momenta again. Therefore, the energy functions may be related as:
\begin{equation}
\begin{array}{rcl}
E\left[ \Gamma^\ast (t) \right] & = & E\left[ \Gamma^\ast (0) \right] + \int_0^t \left[ \dot{Q}_H^\ast + \dot{Q}_C^\ast \right] dt \\

\implies E\left[ \Gamma (0) \right]  & = &  E\left[ \Gamma (t) \right] + \int_0^t \left[ \dot{Q}_H^\ast + \dot{Q}_C^\ast \right] dt \\

\implies -\left[ \dot{Q}_H + \dot{Q}_C \right]  & = &  \left[ \dot{Q}_H^\ast + \dot{Q}_C^\ast \right] \\
\end{array}
\label{eq:appseventeen}
\end{equation}
While writing the last equality, we have used the relation (\ref{eq:appeleven}). In simple terms, the equation (\ref{eq:appseventeen}) says that the heat flow from the thermostats in the time reversed dynamics is exactly equal and opposite to the one in the time-forward dynamics. The generalized work function, therefore, during the time-reversed transition becomes:
\begin{equation}
\begin{array}{rcl}
\Delta W^R_G(\Gamma^\ast(0)) & = & \beta_0 \left( \dfrac{\Delta T_H}{T_H} Q_{H,t}^\ast - \dfrac{\Delta T_C}{T_C} Q_{C,t}^\ast\right) = - \beta_0 \left( \dfrac{\Delta T_H}{T_H} Q_{H,t} - \dfrac{\Delta T_C}{T_C} Q_{C,t} \right) \\
& \implies & \Delta W^R_G(\Gamma^\ast(0))  = -\Delta W^F_G(\Gamma(0)),
\end{array}
\label{eq:appeighteen}
\end{equation}
Since there is no change of $\lambda$ during $A \to SS_1$, the time-reversed trajectory represents the conjugate trajectory moving forward in time. In the same terminology as Evans \textit{et. al.} \cite{No22}, one may therefore, view $\Delta W_G^R(\Gamma^\ast(0))$ synonymously with $\Delta W_G^F(\Gamma^\ast(0))$. Now, we relate the probability of observing a trajectory to its conjugate trajectory:
\begin{equation}
\begin{array}{rcl}
P \left( \Delta W_G^F = -k \right) & = & \int_\Gamma \delta \left( \Delta W_G^F(\Gamma^\ast(0)) + k\right)f_2 \left( \Gamma^\ast(0) \right)d\Gamma^\ast(0) \\ 

& = & \int_\Gamma \delta \left( \Delta W_G^F(\Gamma^\ast(0)) + k\right)f_2 \left( \Gamma(t) \right)d\Gamma(t) \\ 

& = & \int_\Gamma \delta \left( \Delta W_G^F(\Gamma^\ast(0)) + k\right) \exp\left[ -\Delta W_G^F(\Gamma(0)) \right] f_1 \left( \Gamma(0) \right)d\Gamma(0) \\ 

& = &  \int_\Gamma \delta \left( \Delta W_G^F(\Gamma(0)) - k\right) \exp\left[ -\Delta W_G^F(\Gamma(0)) \right] f_1 \left( \Gamma(0) \right)d\Gamma(0) \\ 

& = &  \exp \left[ -k \right] P\left( \Delta W_G^F = k \right)
\end{array}
\label{eq:appnineteen}
\end{equation}
Substituting $\Delta W_G^F$ from equation (\ref{eq:appsixteen}), we get:
\begin{equation}
\begin{array}{rcl}
\dfrac{P \left( \Delta W_G^F = k \right) }{P\left( \Delta W_G^F = -k \right)} & = & \exp \left[ k \right] \\

\dfrac{P \left( \left[ dN_T \dfrac{T_H - T_C}{T_H + T_C} \int_0^t \left[ \eta_C - \eta_H \right]dt \right]_F  = k \right) }{P\left( \left[ dN_T \dfrac{T_H - T_C}{T_H + T_C} \int_0^t \left[ \eta_C - \eta_H \right]dt \right]_F = -k \right)} & = & \exp \left[ k \right] \\

\dfrac{P \left( \left[ \int_0^t \left[ \eta_C - \eta_H \right]dt \right]_F  = k \right) }{P\left( \left[ \int_0^t \left[ \eta_C - \eta_H \right]dt \right]_F = -k \right)} & = & \exp \left[ dN_T \dfrac{T_H - T_C}{T_H + T_C} k \right] \\

\dfrac{P \left( \left[ \bar{\eta}_C - \bar{\eta}_H \right]_F   = k \right) }{P\left( \left[ \bar{\eta}_C - \bar{\eta}_H \right]_F = -k \right)} & = & \exp \left[ dN_T \dfrac{T_H - T_C}{T_H + T_C} kt \right] \\
\end{array}
\label{eq:apptwenty}
\end{equation}
which is exactly what is derived  by Evans \textit{et. al.} in equation (15) of \cite{No22}.

\subsection{Section-III}
In this section we derive equations (8) - (12) of the manuscript. Proceeding analogously like in the previous sections of the supplementary material, but now realizing that during the transition period $\tau$, the work done also features in the first law equation (\ref{eq:appeleven}), we can write:
\begin{equation}
E(\Gamma(t+\tau)) = E(\Gamma(0)) + W + k_B \int_0^{t+\tau} \left[ T_H \langle \Lambda_H \rangle + T_C \langle \Lambda_C \rangle \right] dt^\prime.
\label{eq:apptwentytwo}
\end{equation}
The generalized dimensionless work function now becomes:
\begin{equation}
\exp \left[ \Delta W^F_{G,A \to SS_2}\right] = \left[ \dfrac{\dfrac{\exp \left[ -\beta_0 E(\Gamma(0))\right]}{Z_{\lambda_A}}d\Gamma(0)}{\dfrac{\exp \left[ -\beta_0 E(\Gamma(t+\tau))\right]}{Z_{\lambda_B}}d\Gamma(t+\tau)}\right] \times \dfrac{Z_{\lambda_A}}{Z_{\lambda_B}}.
\label{eq:apptwentythree}
\end{equation}
The partition function at $t+\tau$ is $Z_{\lambda_B}$ because its associated equilibrium state is $B$. The equation gets simplified into:
\begin{equation}
\begin{array}{rcl}
\exp \left[ \Delta W^F_{G,A \to SS_2}\right] & = & \exp \left[ -\beta_0 \left(E(\Gamma(0)) - E(\Gamma(t+\tau)) \right) \right] \times \exp \left[ - \int_0^t\Lambda dt^\prime \right] \\ 

& = & \exp \left[ \beta_0 W\right] \times \exp \left[  \int_0^{t+\tau} \left[ \left( \dfrac{T_H}{T_0} - 1 \right)\Lambda_H + \left( \dfrac{T_C}{T_0} - 1 \right)\Lambda_C \right] dt^\prime \right]  \\ 

& = & \exp \left[ \beta_0 W\right] \times \exp \left[ \dfrac{1}{T_0} \int_0^{t+\tau} \left[ \Delta T_H \Lambda_H - \Delta T_C \Lambda_C \right] dt^\prime \right]  \\ 

& = & \exp \left[ \beta_0 W\right] \times \exp \left[ \beta_0 \int_0^{t+\tau} \left[ \dfrac{\Delta T_H}{T_H} \dot{Q}_H - \dfrac{\Delta T_C}{T_C} \dot{Q}_C  \right] dt^\prime \right]  \\ 
\end{array}
\label{eq:apptwentyfour}
\end{equation}

Therefore, the generalized work function becomes:
\begin{equation}
\exp \left[ \Delta W^F_{G,A \to SS_2}\right] = \beta_0W + \beta_0 \left[ \dfrac{\Delta T_H}{T_H}Q_{H,t+\tau} - \dfrac{\Delta T_C}{T_C}Q_{C,t+\tau}\right],
\label{eq:apptwentyfive}
\end{equation}
which is same as equation (8) of the manuscript. Subtracting equation (\ref{eq:apptwelve}) from equation (\ref{eq:apptwentyfive}) gives the following:
\begin{equation}
\begin{array}{rcl}
\Delta W^F_{G,A \to SS_2} - \Delta W^F_{G,A \to SS_1} & = & \beta_0 W + \beta_0 \left[ \dfrac{\Delta T_H}{T_H}(Q_{H,t+\tau} - Q_{H,t}) - \dfrac{\Delta T_C}{T_C}(Q_{C,t+\tau} - Q_{C,t}) \right] \\
\implies \Delta W^F_{G,A \to SS_2} - \Delta W^F_{G,A \to SS_1} & = & \beta_0 W + \beta_0 \left[ \dfrac{\Delta T_H}{T_H}Q_{H,\tau} - \dfrac{\Delta T_C}{T_C}Q_{C,\tau} \right] \\
\implies \Delta W^F_{G,A \to SS_2} - \Delta W^F_{G,A \to SS_1} & = & \dfrac{f_{eq,A}\left( \Gamma_{SS_1}(0)\right) d\Gamma_{SS_1}(0)Z_{\lambda_A}}{f_{eq,B}\left( \Gamma_{SS_2}(\tau)\right) d\Gamma_{SS_2}(\tau)Z_{\lambda_B}}\\
\end{array}
\label{eq:apptwentysix}
\end{equation}
By looking at the definition of generalized work function, it is evident that the last equality of equation (\ref{eq:apptwentysix}) gives the generalized work function during $SS_1 \to SS_2$ i.e.
\begin{equation}
\begin{array}{rcl}
\Delta W^F_{G,SS_1 \to SS_2} & \equiv & \Delta W^F_{G,A \to SS_2} - \Delta W^F_{G,A \to SS_1}  =  \beta_0 W + \beta_0 \left[ \dfrac{\Delta T_H}{T_H}Q_{H,\tau} - \dfrac{\Delta T_C}{T_C}Q_{C,\tau} \right] \\
& = & \dfrac{f_{eq,A}\left( \Gamma_{SS_1}(0)\right) d\Gamma_{SS_1}(0)Z_{\lambda_A}}{f_{eq,B}\left( \Gamma_{SS_2}(\tau)\right) d\Gamma_{SS_2}(\tau)Z_{\lambda_B}}
\end{array}
\label{eq:apptwentysix_a}
\end{equation}
$f_{eq,A}(\Gamma_{SS_1}(0))d\Gamma_{SS_1}(0)$ represents the probability of the nonequilibrium microstate $\Gamma_{SS_1}(0)$ in the associated equilibrium state $A$. Likewise, $f_{eq,B}(\Gamma_{SS_2}(\tau))d\Gamma_{SS_2}(\tau)$ is the probability in the associated equilibrium state $B$. Equation (\ref{eq:apptwentysix_a}) is the same as the equations (9) and (11) of the manuscript. In a similar manner, by looking at the reverse transition, one can derive the equation (10) of the manuscript.
We will now drop all subscripts except $G$. One can use ergodic consistency -- every microstate in steady states $SS_1$ and $SS_2$ can be obtained from equilibrium states $A$ and $B$, to derive equation (12) of the manuscript:
\begin{equation}
\begin{array}{rcl}
P \left( \Delta W_G^R = -k \right) & = & \int_\Gamma \delta \left( \Delta W_G^R + k\right)f_{eq,B} \left( \Gamma_{SS_2}(\tau) \right)d\Gamma_{SS_2}(\tau) \text{     (by using Liouville's equation)}\\

& = & \int_\Gamma \delta \left( \Delta W_G^R + k\right)\exp \left[ -\Delta W_G^F \right] f_{eq,A} \left( \Gamma_{SS_1}(0) \right) d\Gamma_{SS_1}(0) \dfrac{Z_{\lambda_A}}{Z_{\lambda_B}} \\

& = & \int_\Gamma \delta \left( \Delta W_G^F - k\right)\exp \left[ -\Delta W_G^F \right] f_{eq,A} \left( \Gamma_{SS_1}(0) \right) d\Gamma_{SS_1}(0) \dfrac{Z_{\lambda_A}}{Z_{\lambda_B}}\\

& = & \exp \left[ -k \right] \dfrac{Z_{\lambda_A}}{Z_{\lambda_B}} P \left( \Delta W_G^F = k \right) \\

\end{array}
\end{equation}
Ideally, one should be using the nonequilibrium distributions at time 0 and $\tau$, and not the equilibrium distributions while deriving the previous expression. However, because of a lack of such nonequilibrium distributions, we are limited to using equilibrium distribution functions. As a consequence, the contributions of phase-space compressions, which  seep into the dynamics when nonequilibrium conditions are imposed, cannot be accounted. Therefore, our method works \textit{only} for large $\tau$. Taking $\tau$ large enough, while fixing the time required to reach the steady state, ensures that the contributions arising from the phase-space compressions become negligible.
\bibliography{PRL}
\end{document}